\documentclass[iop,amsmath,amssymb,amsfonts,preprint,superscriptaddress,citeautoscript,byrevtex, onecolumn]{revtex4-1}
\usepackage{graphicx,color}
\usepackage{epstopdf}
\pdfoutput=1 
\usepackage{soul}
\usepackage{ulem}

\begin{document}
\title{ Ballistic spin-transport properties of magnetic tunnel junctions with MnCr-based ferrimagnetic quaternary Heusler alloys }
\author{Tufan Roy}
\email{roy.tufan.a3@tohoku.ac.jp}
\affiliation{Center for Science and Innovation in Spintronics (CSIS), Core Research Cluster (CRC), Tohoku University, Sendai 980-8577, Japan}
\author{Masahito Tsujikawa}
\affiliation{Research Institute of Electrical Communication (RIEC), Tohoku University, Sendai 980-8577, Japan}
\author{Masafumi Shirai}
\affiliation{Center for Science and Innovation in Spintronics (CSIS), Core Research Cluster (CRC), Tohoku University, Sendai 980-8577, Japan}
\affiliation{Research Institute of Electrical Communication (RIEC), Tohoku University, Sendai 980-8577, Japan}
\begin{abstract}
We investigate the suitability of nearly half-metallic ferrimagnetic quaternary Heusler alloys, CoCrMnZ (Z=Al, Ga, Si, Ge) to assess the feasibility as electrode materials of MgO-based magnetic tunnel junctions (MTJ). Low magnetic moments of these alloys originated from the anti-ferromagnetic coupling between Mn and Cr spins ensure a negligible stray field in spintronics devices as well as a lower switching current required to flip their spin direction. We confirmed mechanical stability of these materials from the evaluated values of elastic constants, and 
the absence of any imaginary frequency in their phonon dispersion curves. The influence of swapping disorders on the electronic structures and their relative stability are also discussed. A high spin polarization of the conduction electrons are observed in case of CoCrMnZ/MgO hetrojunctions, independent of terminations at the interface. Based on our ballistic transport calculations, a large coherent tunnelling of the majority-spin $s$-like $\Delta_1$ states can be expected through MgO-barrier. The calculated tunnelling magnetoresistance (TMR) ratios are in the order of 1000\%. A very high Curie temperatures specifically for CoCrMnAl and CoCrMnGa, which are comparable to $bcc$ Co, could also yield a weaker temperature dependece of TMR ratios for CoCrMnAl/MgO/CoCrMnAl (001) and CoCrMnGa/MgO/CoCrMnGa (001) MTJ.

\end{abstract}

\maketitle

\section{Introduction}

Spin-dependent transport properties of electrons have been attracting much attention among the researchers both from the practical applications and basic physics. In spintronics, in addition to the charge current of electrons, the spin currents are utilized in various device applications to reduce the energy consumption. Magnetic tunnel junctions (MTJ) are key elements of spintronic devices, such as non-volatile memories \cite{julliere-pla,miyazaki-jmmm-1995, moodera-prl-1995}. In the MTJ, two magnetic  electrodes are separated by an insulting barrier. If both the magnetic materials have their magnetization direction parallel (antiparallel) to each other, the junction provides a lower (higher) resistance. Tunnelling magnetoresistance (TMR) ratio is proportional to the difference of resistance between the parallel and antiparallel magnetization configurations. A high TMR ratio is desirable in many applications such as magnetoresistive random access memories (MRAM), neuromorphic devices, medical applications, and so on \cite{claude-nature-2007,chen-srep-2017,Lv-apl-2022}.

In the search of suitable electrode materials with a high TMR ratio, different classes of materials have been studied. Heusler alloys are one of these popular classes of materials which are being used as the electrode materials. Many of the Co-based Heusler alloys are reported to show half-metallic electronic structure, for which 100\% spin-polarized conduction electrons are achievable theoretically \cite{miura-prb-2004,troy-prb-2016,galanakis-prb-2002}. Furthermore, the Co-based Heusler alloys show very high Curie temperature ($T_\mathrm {C}$), which is also an essential physical entity to become a potential electrode material. The most studied MTJ with Heusler-alloy electrodes is Co$_2$MnSi/MgO/Co$_2$MnSi \cite{PRL-Hulsen,Miura-PRB-2011,Miura-JPCM-2009,Miura-prb}. It shows the TMR ratio higher than 2,000\% at low temperatures although the TMR ratio diminishes rapidly at room temperature (354\%) \cite{APL-Liu-2012}. The high TMR ratio at lower temperature can be explained as a consequence of coherent tunnelling of $\Delta_1$ electrons through the MgO barrier and a half-metallic electronic structure of Co$_2$MnSi \cite{Butler-PRB,Mathon-prb,Miura-JPCM-2007}.

However, it is indeed a challenge for the researchers to have a weaker temperature dependence of the TMR ratio. To date, $bcc$ Co/MgO/$bcc$ Co shows relatively weak temperature dependence of TMR ratio, owing to very high $T_\mathrm {C}$ \cite{yuasa-apl-2005}. Recently, $bcc$ Co$_3$Mn/MgO/$bcc$ Co$_3$Mn and $bcc$ CoFeMn/MgO/$bcc$ CoFeMn electrodes are also reported to have a lesser temperature dependence of TMR ratio \cite{kunimatsu-apex,jsap-kunimatsu-2019,prapplied-elphic,jalcom-ichinose}.

In magnetization-switching devices such as spin-transfer-torque (STT) MRAM and spin-orbit-torque (SOT) MRAM, switching current, which is proportional to the value of magnetization of free layer, needs to be reduced to reduce the power consumption. Thus ferrimagnetic Heusler alloys could be a promising candidate to investigate. Apart from conventional Co-based ternary Heusler alloys, extensive investigations have been carried out on the quaternary Heusler alloys owing to their diversity and tunability of the electronic and magnetic properties \cite{PRM-Tsuchiya,troy-jmmm,monma-jalcom,onodera-jjap,troy-jphysd}. 
Recently, we investigated Mn-Cr based ferrimagnetic Heusler alloys IrCrMnZ (Z=Al, Ga, Si, Ge) \cite{troy-jphysd}. We have shown that the nearest neighbouring Mn and Cr possesses a strong antiferromagnetic exchange coupling which leads to high $T_\mathrm {C}$. In IrCrMnZ, the replacement of Ir by Co will result in a shorter lattice parameter, leading to a much stronger Mn-Cr exchange coupling, and thus a higher $T_\mathrm{C}$. Another issue with IrCrMnZ is that Ir is an expensive and rare element, and thus there could be a practical obstacle in mass production.

In this study, we theoretically investigate the quaternary Heusler alloys CoCrMnZ (Z=Al, Ga, Si, Ge). These materials are ferrimagnetic with very high $T_\mathrm{C}$, and have reasonable lattice matching with MgO. Previously, CoCrMnAl has been synthesized in its bulk form \cite{prb-enamullah}. However, its heterojunction with MgO has not been studied so far.
In this study, we confirmed the stability of the bulk phases of CoCrMnZ, and then we studied the electronic and magnetic properties of CoCrMnZ/MgO (001) heterojunctions on the basis of the density-functional calculation. Finally, we theoretically investigated the ballistic transport properties of CoCrMnZ/MgO/CoCrMnZ MTJ.

\section{Method}

We carried out structural optimization of bulk CoCrMnZ alloys using Vienna Ab initio simulation package (VASP) \cite{vasp1,vasp2} combined with projector augmented wave (PAW) method \cite{paw}. We used generalized-gradient-approximation (GGA) for the exchange-correlation energy \cite{pbe}. For the Brillouin zone integration, we used a $k$-mesh of 16$\times$16$\times$16. With the optimized structure of bulk CoCrMnZ, we formed a CoCrMnZ(11 ML)/MgO($n$ ML) heterojunctions ($n$=5,7,9) around (001) direction, in which CoCrMnZ unit cell is rotated by 45$^{\circ}$ around the $z$-axis. The atomic positions in the heteojunction were relaxed in the $z$ direction, keeping the in-plane lattice parameter fixed to the equilibrium lattice parameter of bulk CoCrMnZ. During structural optimization, we used a 10$\times$10$\times$1 $k$-mesh, and a denser 16$\times$16$\times$2 $k$-mesh was used for achieving the magnetic moments and density of states. We used a plane-wave cutoff energy of 500 eV in all the calculations using VASP.

To study the dynamical stability of bulk CoCrMnZ, we carried out phonon dispersion calculations. A small displacement method was used for this calculation as implemented in Phonopy code \cite{phonopy1,phonopy2} and interfaced with VASP. Here, we used a 4$\times$4$\times$4 supercell, which contains 256 atoms. The influence of swapping disorder was investigated using a 2$\times$2$\times$2 supercell which includes 128 atoms. 
The chemical disorder of constituent atoms was taken into account by adopting special quasi-random structures (SQS) as implemented in alloy theory automated toolkit package \cite{sqs,atat}.

For the calculations of $T_\mathrm {C}$, we employ mean-field-approximation after evaluating Heisenberg exchange coupling constants using Liechenstein's formalism \cite{licechenstein}. Firstly, we perform self-consistent-field calculations using spin-polarized relativistic Korringa-Kohn-Rostoker package (SPR-KKR) \cite{sprkkr}. We used a full potential mode and the angular momentum expansion of basis function was limited upto $l_{max}$=3. For the integration in the Brillouin zone, we used 917 irreducible $k$-points. All the calculations were performed in scalar-relativistic representation for the valence electrons, where spin-orbit coupling was ignored. The Fermi energy was determined using Llyod's formula \cite{lloyd1, lloyd2}. Note that for the evaluation of $T_\mathrm {C}$ we used both GGA\cite{pbe} as well as local-density-approximation (LDA) \cite{vwn1, vwn2} for the exchange-correlation functional.

Finally, we studied ballistic transport properties of the CoCrMnZ/MgO/CoCrMnZ (001) MTJ using PWCOND code as implemented in Quantum Espresso package \cite{QE1, QE2,pwcond}. For the self-consistent-field calculations, we used a $k$-mesh of 12$\times$12$\times$1, and an energy cutoff of 500 Rydberg for charge density and 50 Rydberg for wave function. The transmission was calculated from the converged charge density and the resultant transmission profile was resolved for a $k$-grid of 100$\times$ 100 for the two-dimensional Brillouin zone in the $k_x$-$k_y$ plane.

\begin{figure}[h]
\includegraphics[width=1.0\textwidth]{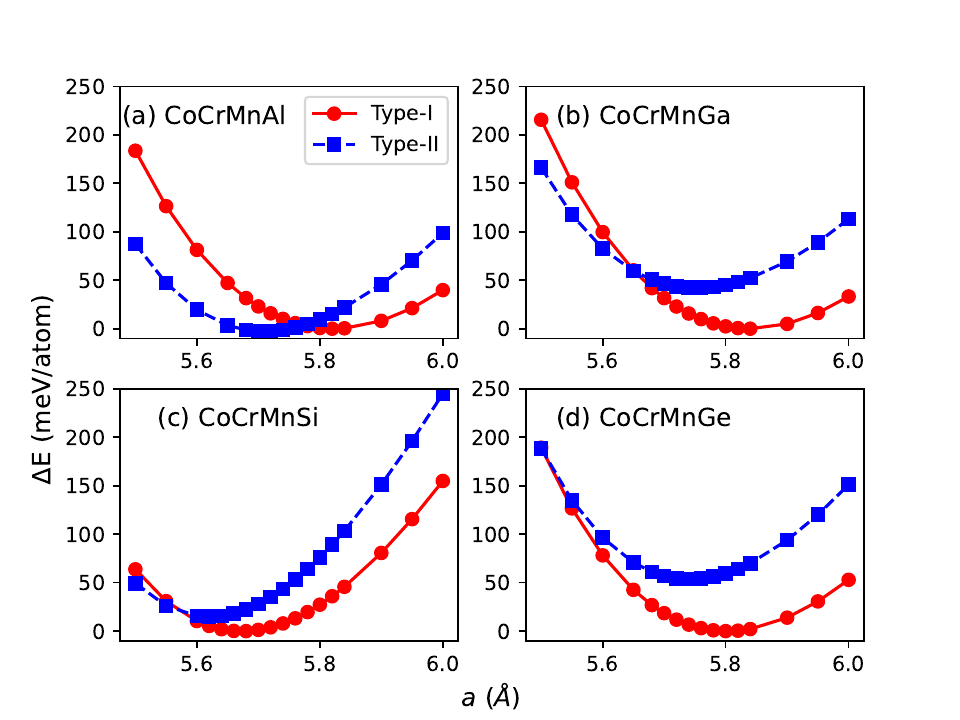}
\caption
{(Color online) Total energy as a function of lattice parameter for (a) CoCrMnAl, (b) CoCrMnGa, (c) CoCrMnSi, and (d) CoCrMnGe, respectively.} 

\end{figure}

\section{Results}
\subsection{Bulk phase}
\subsubsection{Crystal structure}
Quaternary Heusler alloys have a chemical formula of $XX^{\prime}YZ$, where $X$, $X^{\prime}$, and $Y$ are the transition-metal elements and, $Z$ are the main-group $sp$ elements. In the ordered cubic phase, these alloys belong to space group number 216,  $F\bar43m$. There are four atomic sites, $A$ (0, 0, 0), $B$ (0.5, 0.5, 0.5), $C$ (0.25, 0.25, 0.25), and $D$ (0.75, 0.75, 0.75). At first, we considered two different types of configurations as follows\\
Type 1: Cr atom at $B$ site, Mn atom at $C$ site;
Type 2: Mn atom at $B$ site, Cr atom at $C$ site;
in both of these configurations Co and $Z$ atoms occupy $A$ and $D$ sites, respectively.

FIG. 1 shows the total energy ($\Delta E$) of CoCrMnZ as a function of lattice parameter ($a$) for type-I and type-II structures. Note that the origin of the vertical axis is set to the total energy of type-I structure in its equilibrium lattice parameter. We find that except for CoCrMnAl, the other systems have the type-I crystal structure as their lowest energy configuration. However, type-II structure of CoCrMnAl has a slightly lower energy compared to its type-I configuration ($\Delta E$= -2.61 meV/atom). 
Note that the equilibrium lattice parameter of CoCrMnAl in type-II configuration (5.70 \AA) is remarkably shorter than its value in type-I configuration (5.82 \AA). According to the preceding reports \cite{prb-enamullah,Johnson-prb-2016}, the experimental lattice parameter of CoCrMnAl is 5.78 \AA , which is much closer to the type-I configuration. Thus, we consider the type-I structure for all CoCrMnZ (Z=Al, Ga, Si, Ge) hereafter.

\subsubsection{Stability analysis}

\begin{figure}[h]
\includegraphics[width=1.0\textwidth]{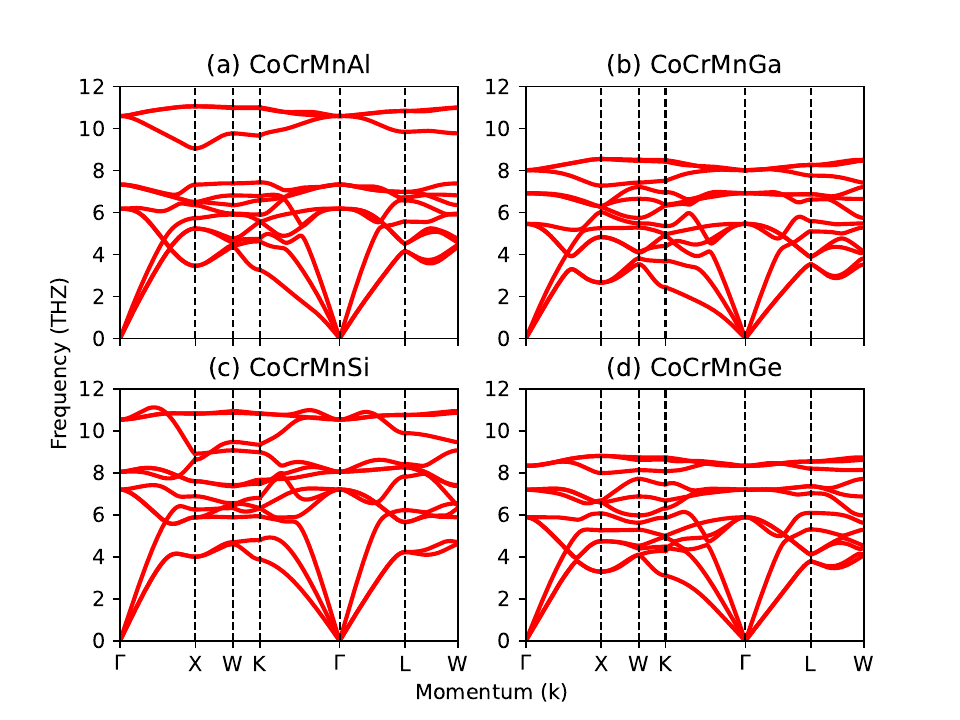}
\caption
{(Color online) Phonon dispersion curves for (a) CoCrMnAl, (b) CoCrMnGa, (c) CoCrMnSi, and (d) CoCrMnGe, respectively.} 

\end{figure}

\textbf{Electronic stability:} In this section we discuss the stability of CoCrMnZ in terms of the formation energy ($E_f$) and the phase separation energy ($\delta E$). The formation energy of a multi-component alloy signifies its stability against decomposition into the equilibrium phases of constituent elements. On the other hand, the phase separation energy of an alloy signifies its stability against decomposition into other alloys and/or elemental phases of constituent atoms. For any energetically stable alloys in the equilibrium conditions, it is necessary that both $E_f$ as well as $\delta E$ are negative. 
We summarize the values of $E_f$ and $\delta E$ in Table 1. We found that for all the cases $E_f$ is negative while $\delta E$ is marginally positive. Possible competing phases are also listed in Table 1. According to Gao \textit{et al.} \cite{Gao-PRM-2019}, a quaternary Heusler alloy with a positive phase separation energy of around 0.1 eV/atom could be synthesized in experiment. For instance, although CoFeCrAl has a phase separation energy of 0.114 eV/atom, this material has been synthesized as thin films by several experimental groups \cite{PRM-Tsuchiya,RRL-Nehra}. Thus we believe that the family of CoCrMnZ alloys could be grown successfully in experiment. Indeed, CoCrMnAl has already been synthesized in bulk \cite{prb-enamullah,Johnson-prb-2016}. Furthermore, in this study we focussed on the spin-dependent transport properties of CoCrMnZ/MgO/CoCrMnZ (001) MTJ, in which CoCrMnZ could be grown as thin films under appropriate non-equilibrium condition.

\begin{table*}[hbt!]
\renewcommand{\thetable}{\arabic{table}}

\centering
\caption{ Formation energy ($E_f$) and phase separation energy ($\delta E$) against competing phases of CoCrMnZ in the unit of eV/atom.}
\begin{tabular}{|c|c|c|c|}
\hline Material & $E_f$   & $\delta E$ & Competing phases\\
\hline CoCrMnAl & -0.19 & +0.11  & AlCo, Cr, Mn  \\
\hline CoCrMnGa & -0.12 & +0.04 & CoGa, Mn$_2$CoGa, Cr\\
\hline CoCrMnSi & -0.33 & +0.06  &  Co$_2$MnSi, Cr$_3$Si, Mn$_3$Si\\
\hline CoCrMnGe & -0.15 & +0.04  & Co$_2$MnGe, Cr$_3$Ge, Mn$_3$Ge\\

\hline
\end{tabular}   
\end{table*}

\textbf{Mechanical stability:} Here, we confirm the mechanical stability of CoCrMnZ, in terms of elastic constants and phonon dispersion curves. For any cubic solid there are three independent elastic constants, $C_{11}$, $C_{12}$, and $C_{44}$. The stability criteria of the cubic phase are as follows: : $C_{11} > 0$,  $C_{44} > 0$, $C_{11} - C_{12} > 0$, and $C_{11} + C_{12} > 0$. As listed in Table 2, all the materials considered here satisfy these criteria. Thus, we conclude that these materials are mechanically stable. We also evaluate bulk and shear moduli, following the Voigt formalism \cite{voigt}. We find that both bulk and shear modulii are also positive for CoCrMnZ and the values are comparable with those for $bcc$ Fe.

The phonon dispersion curves of CoCrMnZ are presented in FIG. 2 to discuss the dynamical stability. For a system, which has an instability against lattice deformation, imaginary frequencies appear in the phonon dispersion curves. For instance, Ni$_2$MnGa and Mn$_2$NiGa, which are well known to undergo a structural transition from its high temperature cubic phase, shows a softening in phonon dispersion curves, implying the instability of the cubic phase \cite{PRB-ener-2012,PRB-kundu-2017}. As found in FIG. 2, all the phonon modes of these materials have real frequencies. Thus, we conclude that the cubic structure of all these materials are dynamically stable.  

\begin{table*}[hbt!]
\renewcommand{\thetable}{\arabic{table}}

\centering
\caption{ Elastic constants of CoCrMnZ. Values for $bcc$ Fe are presented for the sake of comparison.\footnote{Values from previous experiments or calculations \\ 
$^{b}$Ref.\onlinecite{PRL-Caspersen-2004} (theoretical result)
$^{c}$Ref.\onlinecite{PR-Rayne-1961} (experimental result)}}
\begin{tabular}{|c|c|c|c|c|c|}
\hline Material &  $C_{11}$  &   $C_{12}$&   $C_{44}$&Bulk modulus& Shear modulus\\
&(GPa)&(GPa)&(GPa)&(GPa)&(GPa)\\
\hline CoCrMnAl &195 &110 &114 &138 &85\\
\hline CoCrMnGa &189 &120 &109 &143 &79\\
\hline CoCrMnSi &261 &152 &130 &188 &100 \\
\hline CoCrMnGe &211 &131 &112 &157 &83 \\
\hline $bcc$ Fe &277  &132  &89 &180 &82  \\

&271$^{b}$, 243$^{c}$ & 145$^{b}$, 138$^{c}$ &101$^{b}$, 122$^{c}$ &187$^{b}$, 173$^{c}$ &86$^{b}$, 94$^{c}$\\

\hline
\end{tabular}   
\end{table*}

\begin{figure}[h]
\includegraphics[width=1.0\textwidth]{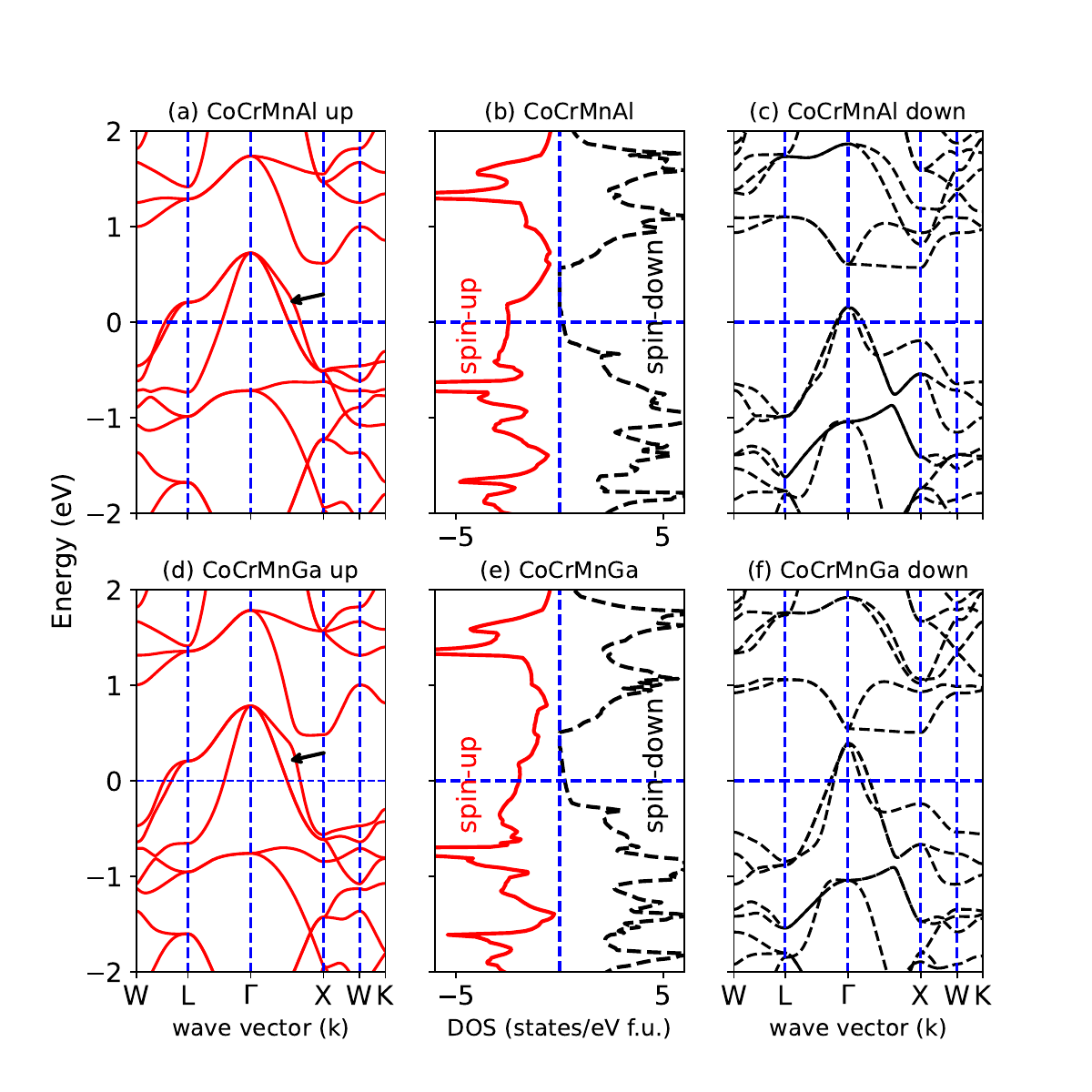}
\caption
{(Color online) Band dispersion curves and density of states for CoCrMnAl (top panels), and CoCrMnGa (bottom panels), respectively. Black arrow in panels (a) and (d) indicates the electronic band with $\Delta_1$ symmetry. } 

\end{figure}

\begin{figure}[h]
\includegraphics[width=1.0\textwidth]{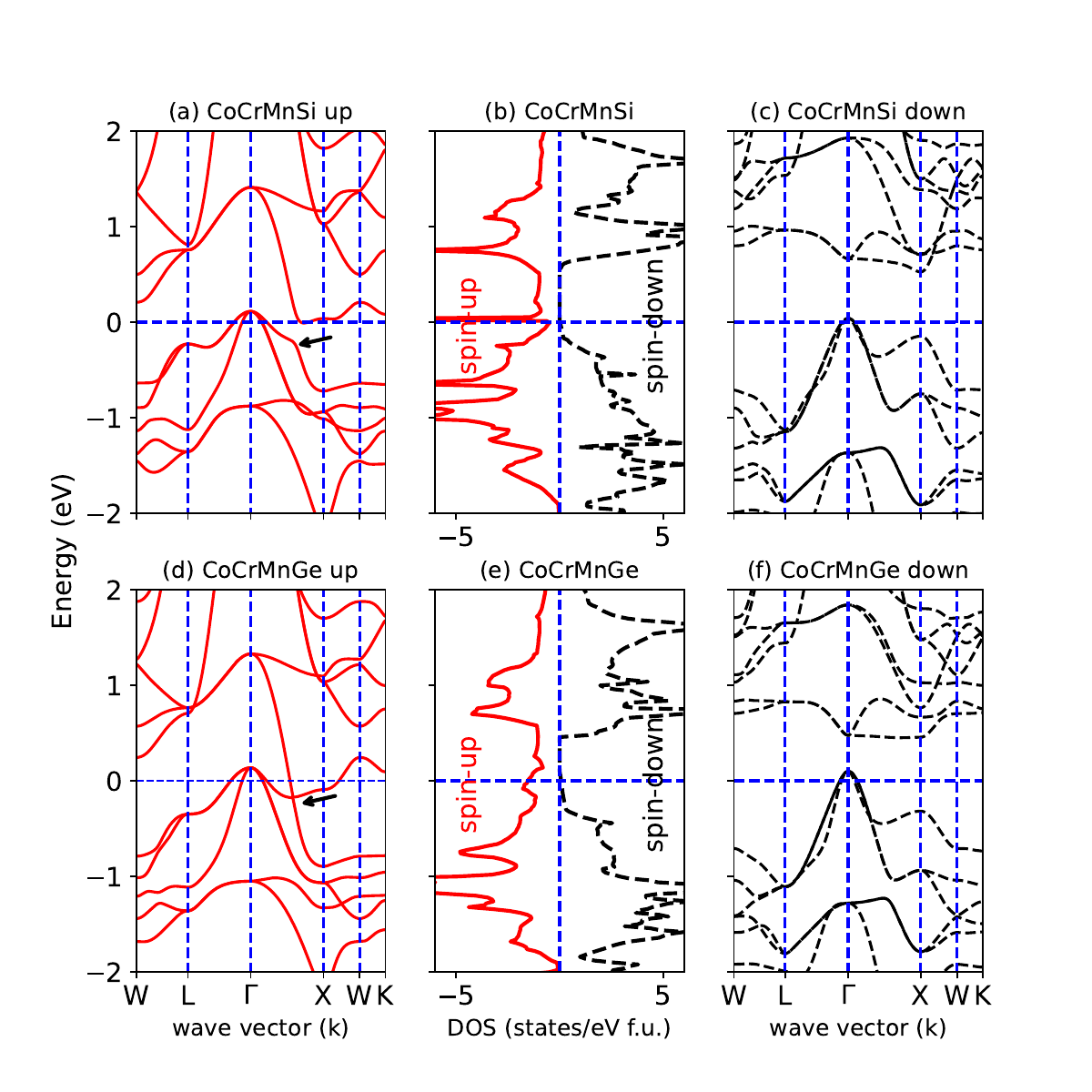}
\caption
{(Color online) Band dispersion curves and density of states for CoCrMnSi (top panels), and CoCrMnGe (bottom panels), respectively. Black arrow in panels (a) and (d) indicates the electronic band with $\Delta_1$ symmetry.} 

\end{figure}

\subsubsection{Electronic structure}

In FIG. 3 we present the electronic band dispersion along high symmetry directions and the density of states (DOS) for CoCrMnAl and CoCrMnGa. For CoCrMnAl and CoCrMnGa the DOS at the Fermi level ($E_\mathrm{F}$) in the majority-spin channel are 2.47 and 1.92 states/eV per formula unit, respectively. On the other hand, there are hole pockets at the top of the valence band around $\Gamma$ point in the minority-spin channel, leading to smaller DOS at $E_\mathrm{F}$, i.e. 0.16 and 0.25 states/eV per formula unit for CoCrMnAl and CoCrMnGa, respectively. As a result, the values of spin polarization are 88\% and 77\%, respectively, for CoCrMnAl and CoCrMnGa.
FIG. 4 depicts the electronic band dispersion and the DOS of CoCrMnSi and CoCrMnGe. These two materials also show metallic behaviour for the majority-spin channel, although the DOS at $E_\mathrm{F}$ are much smaller compared to those of CoCrMnAl and CoCrMnGa. The values of the majority-spin DOS at $E_\mathrm{F}$ are 0.6 and 1.6 states/eV per formula unit for CoCrMnSi and CoCrMnGe, respectively. However, the minority-spin DOS almost vanishes for both CoCrMnSi and CoCrMnGe, leading to a very high spin polarization of the conducting electrons, which are approximately 100\% for CoCrMnSi and 92\% for CoCrMnGe.

Both in FIG. 3 and 4, we marked the totally symmetric $\Delta_1$ band by an black arrow in the majority-spin channel along the $\Gamma$-X
direction ($\Delta$ line). The $\Delta_1$ band is predominantly composed of $s$, $p_z$, and $d_{z^2}$ orbitals of Co, Cr, and Mn atoms. It has been reported that in MgO-based MTJ the $\Delta_1$ electrons have the slowest decay rate through the MgO barrier, which originates from the complex band structure of MgO. Thus, the presence of $\Delta_1$ electrons in the majority-spin channel plays a pivotal role in obtaining large transmission and low resistance-area (RA) product in the parallel magnetization configuration of the MTJ.

\subsection {Magnetic properties}

CoCrMnZ (Z=Al, Ga, Si, Ge) have the ferrimagnetic ground state, in which the spin-moments of Cr atoms have the opposite directions with respect to Co and Mn spin-moments. A strong antiferromagnetic exchange coupling between the nearest neighbouring Cr and Mn spins stabilizes the ferrimagnetic ground state. It is to be noted that the total magnetic moments of half-metallic full Heusler alloys obey Slater-Pauling rule \cite{galanakis-prb-2002} i.e. the total magnetic moment ($\mu_{total}$) is related to the valence electron number ($N_v$) as follows:
\begin{equation}
\mu_{total} = N_v -24    
\end{equation}
In CoCrMnAl and CoCrMnGa, the value of $N_v$ is 25, and it becomes 26 for CoCrMnSi and CoCrMnGe. A slight deviation from the Slater-Pauling rule found in CoCrMnZ results in the reduction of spin polarization from its ideal value of 100\%. 

\begin{figure}[h]
\includegraphics[width=0.8\textwidth]{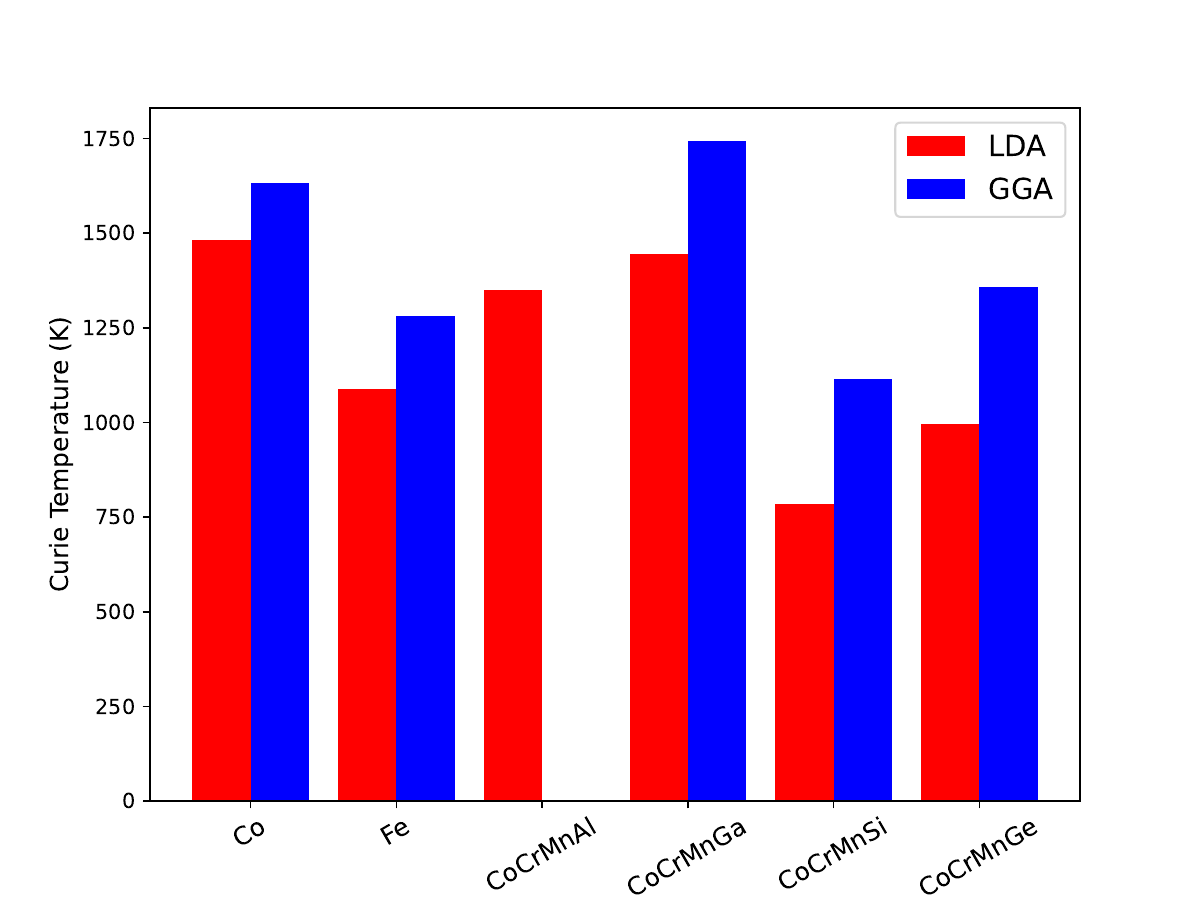}
\caption
{(Color online) Curie temperatures for (a) CoCrMnAl, (b) CoCrMnGa, (c) CoCrMnSi, and (d) CoCrMnGe, respectively, evaluated by using LDA and GGA . Those for $bcc$ Co and $bcc$ Fe are also presented for the sake of comparison. ($T_\mathrm {C}$ of CoCrMnAl did not converge using GGA).} 

\end{figure}

\begin{table*}[hbt!]
\renewcommand{\thetable}{\arabic{table}}

\centering
\caption{ Calculated magnetic moments of CoCrMnZ and their equilibrium lattice parameter ($a_{opt}$).}
\begin{tabular}{|c|c|c|c|c|c|}
\hline Material & $a_{opt}$  & $\mu_{total}$ & $\mu_{\mathrm Co}$ & $\mu_{\mathrm Cr}$ & $\mu_\mathrm {Mn}$\\
&(\AA)&($\mu_{\mathrm B}$)&($\mu_{\mathrm B}$)&($\mu_{\mathrm B}$)&($\mu_{\mathrm B}$)\\
\hline CoCrMnAl & 5.822 &1.02 &0.67& -2.39 & 2.80 \\
\hline CoCrMnGa & 5.840 &1.08 & 0.70& -2.49& 2.90 \\
\hline CoCrMnSi & 5.672 &1.99 & 0.77&-1.42 & 2.63 \\
\hline CoCrMnGe & 5.803 &2.01 & 0.85& -1.79 & 2.90 \\

\hline
\end{tabular}   
\end{table*}

High TMR ratio in the MTJ with the MgO barrier is generally ensured if the electrode material has partially occupied $\Delta_1$ band only in one spin channel owing to the coherent tunnelling \cite{Butler-PRB, Mathon-prb}. However, the TMR ratio at the room temperature is remarkably reduced from its low temperature value because of thermal fluctuation of spin moments at the interfacial region with MgO \cite{Miura-PRB-2011}. One way to suppress the thermal fluctuation of the interfacial moments is to choose electrode materials with very high $T_\mathrm {C}$. For example, metastable $bcc$ Co is known to have $T_\mathrm {C}$ $\approx$ 1500 K and the MTJ with bcc Co electrodes has a very weak temperature dependence of TMR ratio \cite{yuasa-apl-2005}. The values of TMR ratios at low and room temperatures are 507\%  and 410\%, respectively \cite{yuasa-apl-2005}. In this study, we have calculated the $T_\mathrm {C}$ within mean-field-approximation (MFA). It is well-known that MFA generally overestimates the $T_\mathrm {C}$ as it ignores the short-range spin correlation in the paramagnetic phase. A study by Zagrebin \textit{et al.} shows that the values of $T_\mathrm {C}$ evaluated within MFA in combined with LDA are in better agreement with the experimental values \cite{zagrebin-2019}. In FIG. 5, we plot the values of $T_\mathrm {C}$ evaluated from the electronic structure calculated with LDA and GGA. (Note that $T_\mathrm {C}$ of CoCrMnAl did not converge for GGA). Here, we also include $bcc$ Fe and $bcc$ Co for the sake of comparison in addition to CoCrMnZ. CoCrMnSi and CoCrMnGe have $T_\mathrm {C}$ values almost half that of the value of $bcc$ Co. However, CoCrMnAl and CoCrMnGa have the $T_\mathrm {C}$ values close to $bcc$ Co. These findings show that the MTJ with CoCrMnAl and CoCrMnGa could have a weaker temperature dependence of TMR ratio.

\begin{figure}[h]
\includegraphics[width=1.0\textwidth]{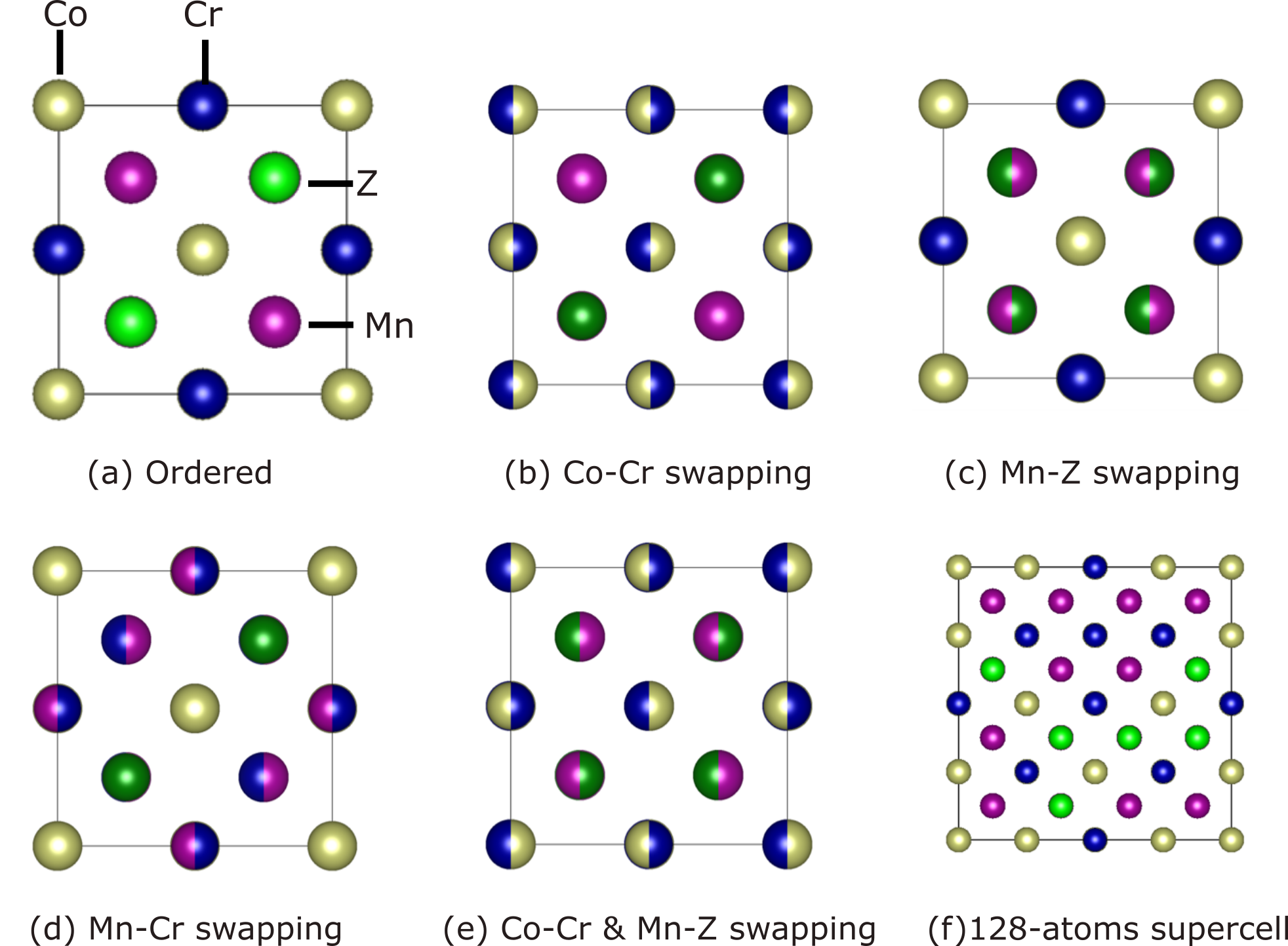}
\caption
{(Color online) Schematic representation of various atomic ordering considered: (a) Ordered, $Y$ , (b) Co-Cr swapping, $L2_1$-I, (c) Mn-Z swapping, $L2_1$-II, (d) Mn-Cr swapping, $XA$, and (e) Co-Cr and Mn-Z swapping, $B2$. (f) The atomic configuration in the SQS for the $B2$ phase.} 

\end{figure}

\begin{figure}[h]
\includegraphics[width=1.0\textwidth]{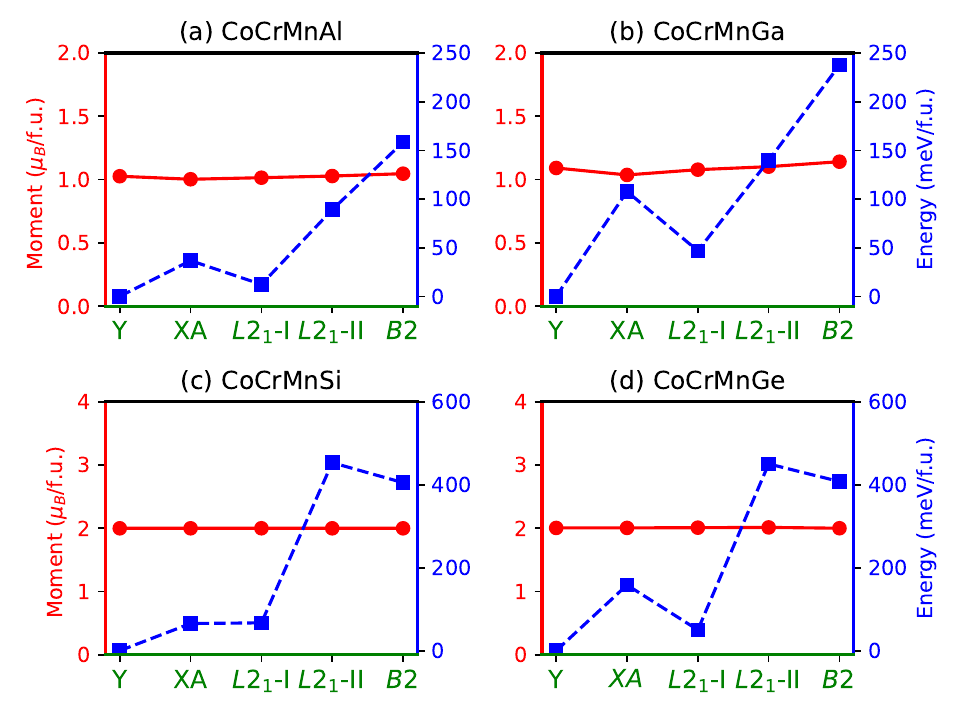}
\caption
{(Color online) Total magnetic moment (red circles) and total energy (blue squares) of ordered Y and disordered $XA$, $L2_1$-I, $L2_1$-II, and $B2$ phases for (a) CoCrMnAl, (b) CoCrMnGa, (c) CoCrMnSi, and (d) CoCrMnGe, respectively.} 

\end{figure}

\begin{figure}[h]
\includegraphics[width=1.0\textwidth]{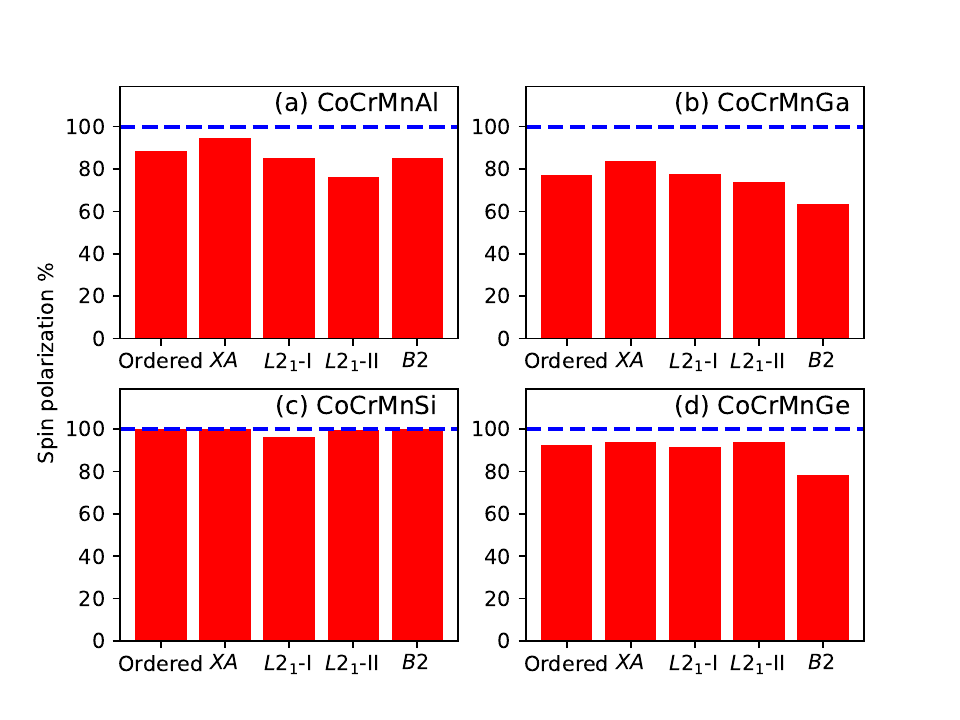}
\caption
{(Color online) Spin polarization at $E_\mathrm{F}$ of ordered Y and disordered $XA$, $L2_1$-I, $L2_1$-II, and $B2$ phases for (a) CoCrMnAl, (b) CoCrMnGa, (c) CoCrMnSi, and (d) CoCrMnGe, respectively.} 

\end{figure}

\subsubsection{Influence of swapping disorders}

In case of quaternary Heusler alloys, it is one of the major challenges to synthesize ordered $Y$-structure. Various types of antisite or swapping disorders may form during its experimental fabrication. Till now our discussion was limited to the ordered $Y$-structure. In this section we discuss possibility of the formation of disordered phases in terms of their total energies with respect to the ordered phase. We also discuss the influence of various types of swapping disorder on the electronic structure. Here, we have considered four different types of disordered phases, such as (i) Co-Cr swapping disorder ($L2_1$-I), (ii) Mn-Z swapping disorder ($L2_1$-II), (iii) Mn-Cr swapping disorder ($XA$), and (iv) Co-Cr and Mn-Z swapping disorder ($B2$), as depicted in FIG. 6.

FIG. 7 compares the relative energy of the disordered phases with respect to the ordered one. We find that the ordered $Y$-structure is the energetically most stable, followed by $L2_1$-I, and $XA$ types of disorder. The swapping disorders between the transition-metal elements themselves tend to be energetically less expensive, hence more likely to occur. It can be understood that all three transition elements namely, Co, Cr and Mn have very close atomic numbers, atomic radii, and valence electron numbers, which could promote intermixing between themselves. On the other hand, $L2_1$-II or $B2$ disorder which involves swapping between main-group elements and the transition-metal elements, are energetically unfavourable, hence more unlikely to occur.
FIG. 7 also depicts the influence of various disorder on the total magnetic moment. It is clearly observed that the total magnetic moment is mostly robust against the swapping disorders.

As the spin polarization of conduction electrons is an important physical quantity for understanding spin-dependent transport properties, we also show the influence of swapping disorder on the spin polarization in FIG. 8. All the materials maintain high spin polarization in the disordered phases. For CoCrMnSi, in particular, the spin polarization is almost 100\% even in the disordered phases. Thus, we conclude that the swapping disorder hardly affect spin polarization significantly.

\subsection{Interface with MgO}

\begin{figure}[h]
\includegraphics[width=1.0\textwidth]{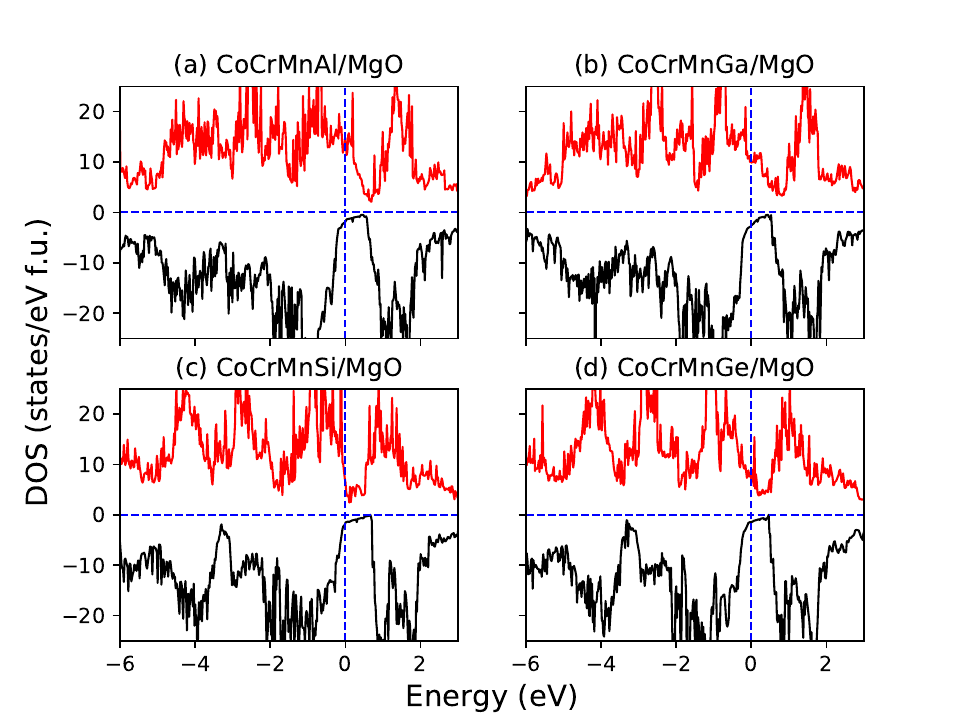}
\caption
{(Color online) Density of states (DOS) of CoCr-terminated (a) CoCrMnAl/MgO, (b) CoCrMnGa/MgO, (c) CoCrMnSi/MgO, and (d) CoCrMnGe/MgO heterojunctions, respectively.} 

\end{figure}
\begin{figure}[h]
\includegraphics[width=1.0\textwidth]{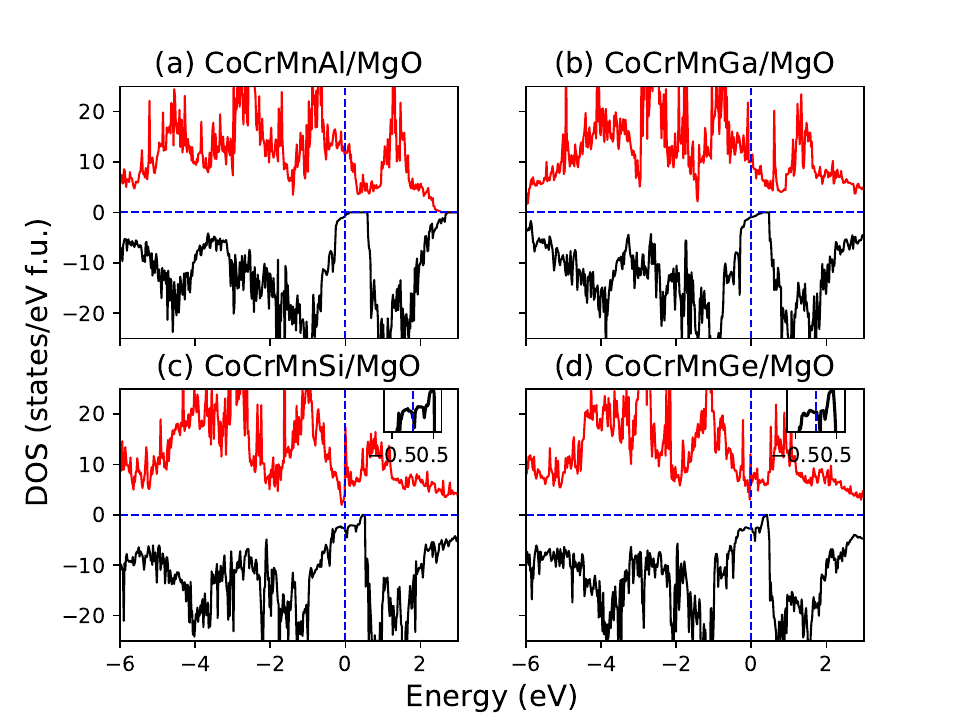}
\caption
{(Color online) Density of states (DOS) of MnZ-terminated (a) CoCrMnAl/MgO, (b) CoCrMnGa/MgO, (c) CoCrMnSi/MgO, and (d) CoCrMnGe/MgO heterojunctions, respectively. Insets of (c) and (d) show the DOS in the vicinity of $E_\mathrm{F}$ for the minority-spin channel.} 

\end{figure}

\subsubsection{Electronic structure and stability consideration}

In this section we discuss the electronic structure of the CoCrMnZ/MgO (001) heterojunctions. For CoCrMnZ, we considered two types of terminations, namely CoCr-terminated and MnZ-terminated interfaces. In both the cases the atoms on the terminated layer of the CoCrMnZ are located at the top of the O-atom of MgO. The interfacial lattice mismatch between CoCrMnZ Heusler alloys ($a_{opt}/\sqrt{2}$) and MgO (4.21 \AA) are 2.2\%, 1.9\%, 4.7\%, and 2.5\%, for CoCrMnAl, CoCrMnGa, CoCrMnSi, and CoCrMnGe, respectively.

FIG. 9 and 10 show the DOS of CoCrMnZ/MgO (001) heterojuntions for the CoCr- and MnZ-terminated interfaces, respectively. For both the terminations high spin polarization in their respective bulk phases have been maintained. The values of the spin polarization in the CoCr-terminated interfaces are 77\%, 59\%, 61\%, and 64\%, respectively for CoCrMnAl/MgO, CoCrMnGa/MgO, CoCrMnSi/MgO, and CoCrMnGe/MgO. For the MnZ-terminated interfaces the values of the spin polarization become 90\%, 83\%, 67\% and 41\% for CoCrMnAl/MgO, CoCrMnGa/MgO, CoCrMnSi/MgO, CoCrMnGe/MgO, respectively. It is observed that the in-gap states in the minority-spin channel appear at $E_\mathrm {F}$ in the MnSi-terminated CoCrMnSi/MgO and MnGe-terminated CoCrMnGe/MgO heterojunctions. A further analysis shows that these in-gap states mainly arise from the interfacial Si (Ge) atoms because of their weak bonding with the nearest neighbouring O atoms. This observation is quite consistent with our earlier studies on the related systems \cite{Miura-prb,troy-jphysd,troy-jpcm}.

Furthermore, we compare the relative stability between the CoCr- and MnZ-terminated heterojunctions using the formalism adopted in the previous studies \cite{troy-jphysd}. We calculate the formation energy ($E_{form}$) from the following formula:
\begin{equation}
E_{form} = E_{Total}- \sum_i N_i E_i 
\end{equation}
where $E_{Total}$ is the total energy of the relevant heterojunction, $N_i$ is the number of each constituent element, and $E_i$ is the total energy of each element in their ground state. According to the $E_{form}$ evaluated for both the terminations, the MnZ-terminated interface is energetically favourable compared to the CoCr-terminated one. The difference in the formation energies are 1.99 eV, 1.68 eV, 1.31 eV, and 1.27 eV, respectively for CoCrMnAl/MgO, CoCrMnGa/MgO, CoCrMnSi/MgO, and CoCrMnGe/MgO heterojunctions.   

As the MnZ-terminated heterojunctions are more stable compared to the CoCr-terminated ones, thus we focus on the MnZ-terminated interface in the following study of  spin-transport properties of CoCrMnZ/MgO/CoCrMnZ (001) MTJ.

\subsubsection{Ballistic transport properties}

\begin{figure}[h]
\includegraphics[width=0.6\textwidth]{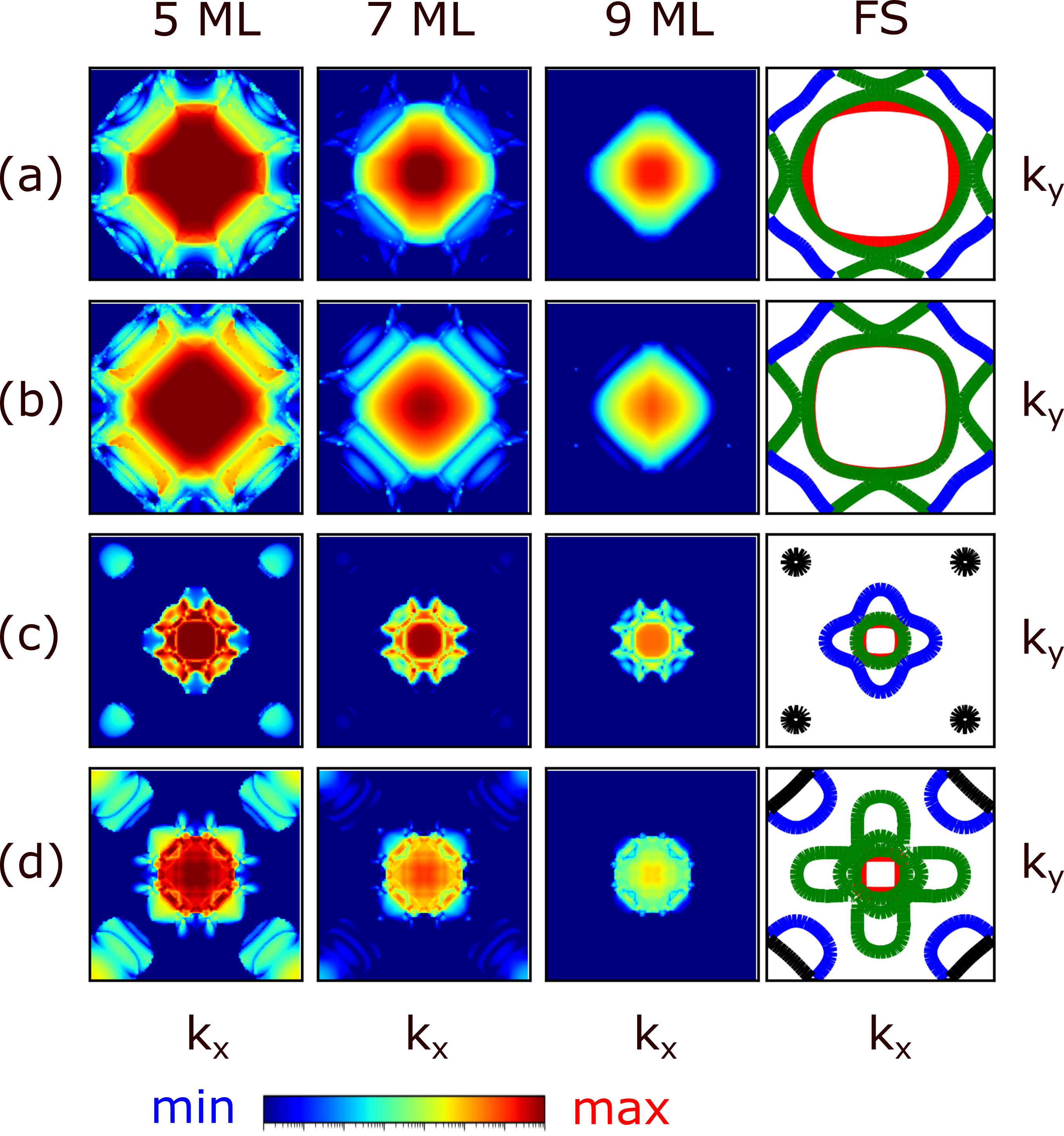}
\caption
{(Color online) In-plane wavevector ($k_x$, $k_y$) dependence of the majority-spin transmittance at $E_\mathrm {F}$ in the parallel magnetization configuration for (a) CoCrMnAl/MgO($n$ ML)/CoCrMnAl (001), (b) CoCrMnGa/MgO($n$ ML)/CoCrMnGa (001), (c) CoCrMnSi/MgO($n$ ML)/CoCrMnSi (001), and (d) CoCrMnGe/MgO($n$ ML)/CoCrMnGe (001) MTJ, respectively, ($n$=5,7,9). The rightmost panels show the cross-section of the majority-spin Fermi surface plotted using FermiSurfer\cite{fermisurfer}.}

\end{figure}

\begin{figure}[h]
\includegraphics[width=0.6\textwidth]{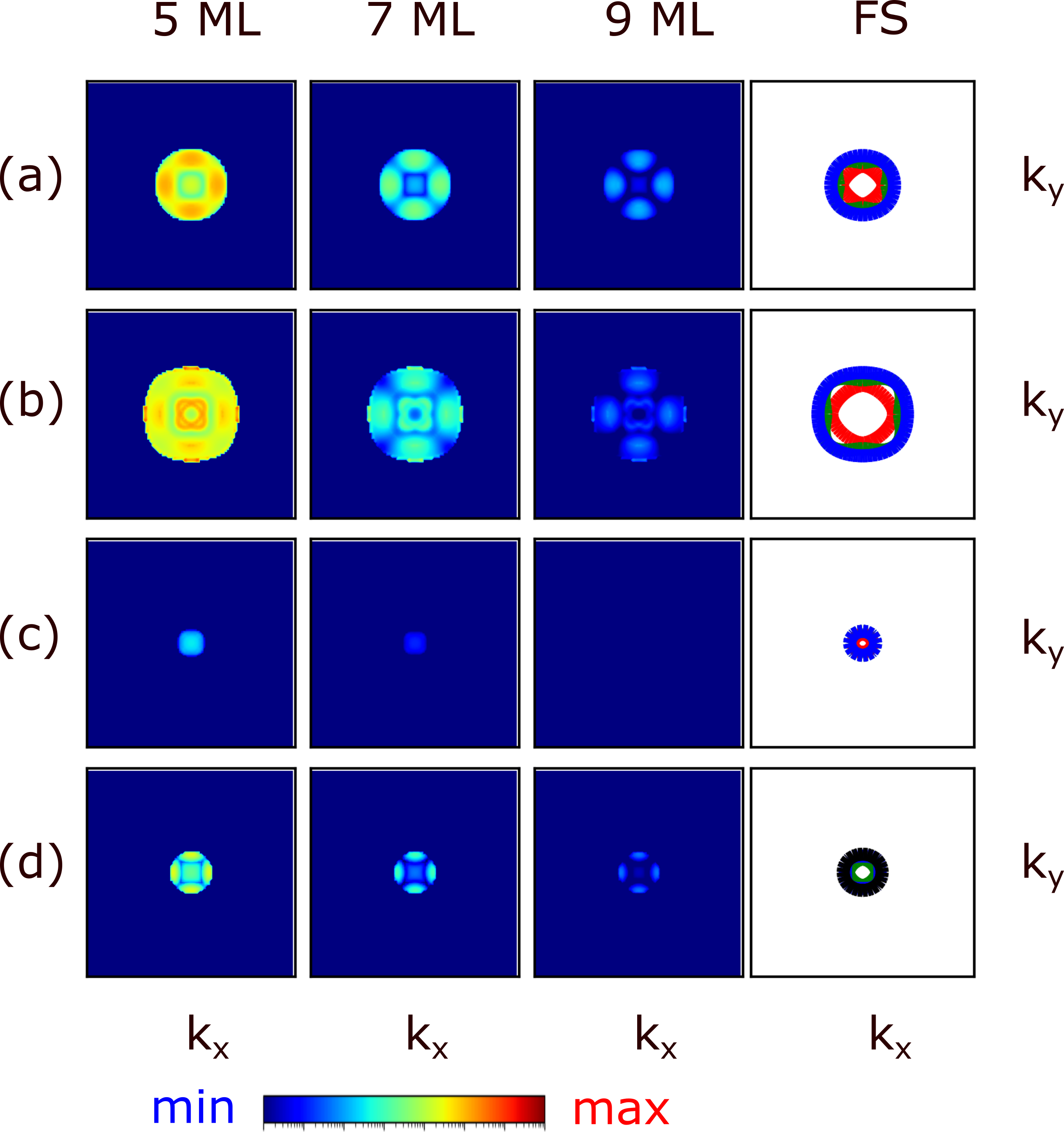}
\caption
{(Color online) In-plane wavevector ($k_x$, $k_y$) dependence of the minority-spin transmittance at $E_\mathrm {F}$ in the parallel magnetization configuration for (a) CoCrMnAl/MgO($n$ ML)/CoCrMnAl (001), (b) CoCrMnGa/MgO($n$ ML)/CoCrMnGa (001), (c) CoCrMnSi/MgO($n$ ML)/CoCrMnSi (001), and (d) CoCrMnGe/MgO($n$ ML)/CoCrMnGe (001) MTJ, respectively, ($n$=5,7,9). The rightmost panels show the cross-section of the minority-spin Fermi surface plotted using FermiSurfer\cite{fermisurfer}.}

\end{figure}

\begin{figure}[h]
\includegraphics[width=0.6\textwidth]{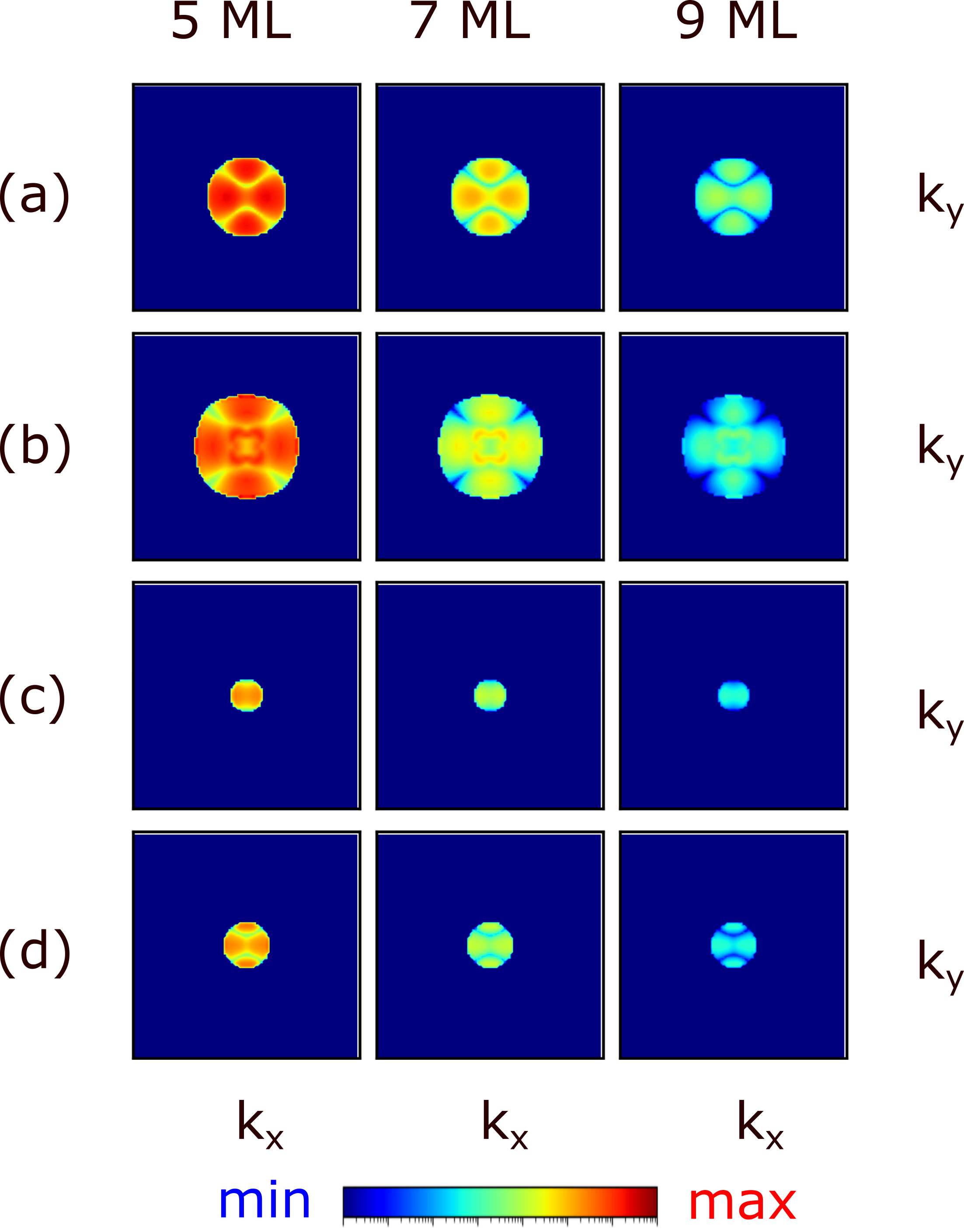}
\caption
{(Color online) In-plane wavevector ($k_x$, $k_y$) dependence of the majority-spin transmittance at $E_\mathrm {F}$ in the antiparallel magnetization configuration for (a) CoCrMnAl/MgO($n$ ML)/CoCrMnAl (001), (b) CoCrMnGa/MgO($n$ ML)/CoCrMnGa (001), (c) CoCrMnSi/MgO($n$ ML)/CoCrMnSi (001), and (d) CoCrMnGe/MgO($n$ ML)/CoCrMnGe (001) MTJ, respectively, ($n$=5,7,9).}
\end{figure}

In FIG. 11, we show the in-plane wavevector ($k_x$, $k_y$) dependence of the electron transmission profile of CoCrMnZ/MgO ($n$ ML)/CoCrMnZ (001) MTJ in the parallel magnetization configuration, where $n$=5, 7, 9. In almost all the cases, a large transmission is centred at ($k_x$, $k_y$) = (0, 0) of the two-dimensional Brillouin zone. This feature can be assigned to the presence of partially occupied $\Delta_1$ band at the majority-spin channel. We also found that as the MgO layers become thicker, off-centre transmission vanishes, except for the transmission around ($k_x$, $k_y$) = (0, 0). To understand the transmission profile for the MTJ with each electrode material, we show the cross-section of Fermi surface (FS) for the tetragonal unit cell of CoCrMnZ on a plane perpendicular to the  (001) direction and passing through the $\Gamma$ point. A clear correlation can be seen between the shape of the FS cross-section and the transmission profile through the MgO barrier, specifically when the MgO layers are thinner.

The minority-spin transmission probability has been presented in FIG. 12, as a function of the in-plane wavevector. The minority-spin transmittance is limited only close to ($k_x$, $k_y$)=(0,0). We have already observed in the band structure of CoCrMnZ that the top of the valence band just touches $E_\mathrm {F}$ near the $\Gamma$ point. It leads to a smaller FS for the minority-spin channel, as shown in the rightmost panels of FIG. 12, which is consistent with a narrow conduction channel for the minority-spin electrons.

Now, to discuss the spin-filtering properties of the CoCrMnZ/MgO/CoCrMnZ MTJ, the MgO-thickness dependence of the transmission profile of electrons in the antiparallel magnetization configuration is presented in FIG. 13. The transmission in the antiparallel configuration involves the tunnelling of electrons from the majority-spin channel of one electrode to the minority-spin channel of the other. Here, we study symmetric MTJ, where both the electrodes on the opposite sides of barrier are identical. Thus, the transmission in the antiparallel configuration is related to availability of both the minority- and majority-spin electronic states at $E_\mathrm {F}$. In CoCrMnZ electrodes, we found a smaller FS for the minority-spin channel, thus the transmission is also limited to the region around ($k_x$, $k_y$)=(0,0). As anticipated, a small value of transmittance in antiparallel magnetization configuration is obtained here, which is highly required to have a higher TMR ratio of the MTJ. We evaluate the TMR ratio for the MTJ with 7ML MgO barrier, which is almost equivalent to 1.5 nm of barrier thickness. The values of TMR ratios are quite huge, i.e. 8,631\%, 4,597\%, 70,295\%, and 5,804\% for the CoCrMnAl/MgO(7 ML)/CoCrMnAl, CoCrMnGa/MgO(7 ML)/CoCrMnGa, CoCrMnGa/MgO(7 ML)/CoCrMnGa, CoCrMnGe/MgO(7 ML)/CoCrMnGe MTJ, respectively.

\section{Summary}

By using the first-principles calculations, we found that CoCrMnZ (Z=Al, Ga, Si, Ge) have the ferrimgnetic ground state with very high $T_\mathrm{C}$. In the respective bulk phases, they have very high spin polarization of conduction electrons, which is robust against swapping disorder. Although these systems possess a slight positive phase separation energy, the negative formation energy, and the absence of imaginary frequency in the phonon dispersion curves ensure the possibility of growing these materials as thin films. We also report here that a large transmission of the majority-spin electrons in the CoCrMnZ/MgO/CoCrMnZ MTJ in the parallel magnetization configuration owing to the presence of $\Delta_1$ band across $E_\mathrm{F}$, whereas the transmission in the antiparallel magnetization configuration is remarkably low. It leads to a colossal TMR ratio much higher than 1000\%. Thus, we conclude that the high spin polarization of conduction electrons, the good lattice matching with MgO, and a huge TMR ratio along with very high $T_\mathrm{C}$, specifically for CoCrMnAl and CoCrMnGa, could make them technologically important in spintronics device applications, which awaits experimental validation.

\section{Acknowledgements} This work was partially supported by JST CREST (No.
JPMJCR17J5) and by CSIS, Tohoku University. The authors thank S. Mizukami and A. Hirohata for fruitful discussions.

{}

\clearpage

\end{document}